# MULTI-DIMENSIONAL PASSWORD GENERATION TECHNIQUE FOR ACCESSING CLOUD SERVICES

Dinesha H A[1] and Dr.V.K Agrawal[2]

[1]Assistant Professor, Department of ISE & CORI, PES Institute of Technology, Bangalore, India
`sridini@gmail.com`

[2]Professor & Director, Dept of ISE & CORI, PES Institute of Technology, Bangalore, India
`vk.agrawal@pes.edu`

## ABSTRACT

*Cloud computing is drastically growing technology which provides an on-demand software, hardware, infrastructure and data storage as services. This technology is used worldwide to improve the business infrastructure and performance. However, to utilize these services by intended customer, it is necessary to have strong password authentication. At present, cloud password authentication can be done in several ways, such as, textual password, graphical and 3D password. In this paper, we are proposing the strong password generation technique by considering multiple input parameters of cloud paradigm referred as a multidimensional password. This paper presents the multidimensional password generation technique along with architecture, sequence diagrams, algorithms and typical user interfaces. At the end, we derive the probability of breaking our authentication system.*

## 1. INTRODUCTION

CLOUD computing technology is a service-based, Internet-centric, safe, convenient data storage and network computing service [1]. It is an internet-based model for enabling a convenient and on-demand network access to a shared pool of configurable computing resources [2]. The provision of various services over internet is possible through this technology which connotes software, hardware, data storage and infrastructure. Cloud computing service provider delivers the applications via internet. These services are accessed from web browsers, desktop and mobile apps. Cloud Computing Technologies are grouped into 4 sections which include SaaS, DSaaS, IaaS and PaaS [3] [4]. SaaS (Software as a Service) is an on-demand application service. It delivers software as a service over the Internet. It eliminates the need of installing and running the application on the customer's own computers [3] [4]. PaaS (Platform as a Service) is an on-demand platform service to host customer application. DSaaS (Data Storage as Services) is an on-demand storage service. IaaS (Infrastructure as a Service) is an on- demand infrastructure service. It delivers the computer infrastructure – typically a platform virtualization environment – as a service, along with raw (block) storage and networking.

In order to make a secure usage of the services provided by the cloud, cloud authentication systems can use different password techniques like: i) Simple text password ii) Graphical password and iii) 3D password object. But each of this has its own drawbacks. The weakness of textual password authentication system is that, it is easy to break and it is very much vulnerable to dictionary or brute force attacks. Graphical passwords have memory space which is found to be less than or equal to the textual password space. But since this particular technique is based on the idea that a picture speaks thousand words, it preferred over the previous technique. However, some of the graphical password schemes require a long time to be performed [5] [6]. Thus, they are also constrained by time complexity. Similarly even 3D- password authentication has its own limitations [7]. Another simple approach is to generate and authenticate the multi-dimensional password by considering many aspects of cloud paradigm. The multi-dimensional password gets generated by considering many aspects and inputs such as, logos, images, textual information and signatures etc. By doing so, the probability of brute force attack for breaking the password can be reduced to a large extent. Hence it has motivated us to introduce a multi-dimensional password generation/authentication technique in secure cloud transmission by ensuring strong passwords.

This paper is organized in the following manner: In section 2, we have proposed multi-dimensional password generation technique along with the architecture and sequence diagram. In section 3, we present the detailed design of proposed multi-dimensional password authentication/generation technique. Section 4 concludes this paper along with the future work.

## 2. PROPOSED MULTI-DIMENSIONAL PASSWORD GENERATION TECHNIQUE

According to this technique, access to the cloud is authenticated using a multi-dimensional password. It generates the multi-dimensional password by considering the many parameter of cloud paradigm such as: vendor details, consumer details, services, privileges and etc. These parameters considered as input dimension. These many dimensions (input) combined together and produces multidimensional password. Fig. 1 depicts the architecture diagram of multi-dimensional authentication system. This has two separate entities i) cloud service provider which provides variety of cloud services and ii) Authenticated client organization to use cloud services (Before using cloud services, company authentication confirms with service agreement from cloud vendors). This architecture helps in checking authentication against the services and privileges. The multi-dimensional password is generated by considering many aspects and confidential inputs such as logos, images, textual information and signatures etc. This is portrayed in figure 2. With the help of this technique, the probability of brute force attack for breaking the password is greatly reduced. Fig 3 shows the sequence diagram of multi-dimensional password generation system.

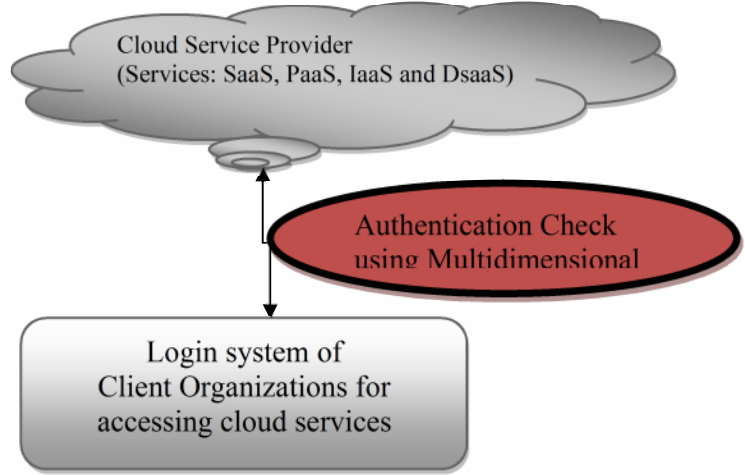

Fig 1: Architecture of multi-dimensional authentication system

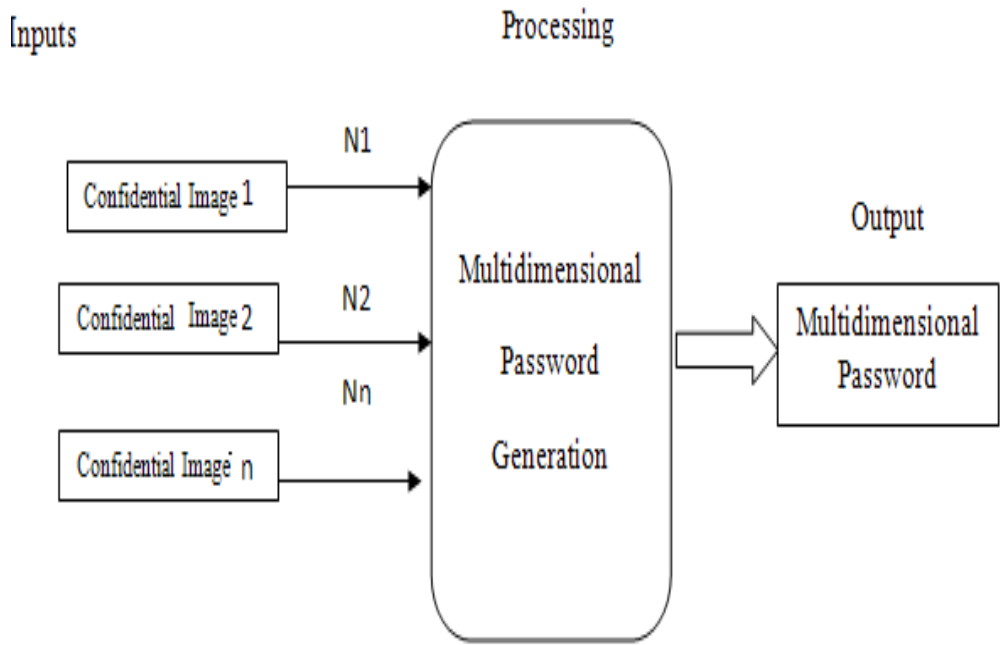

Fig 2: Multi-dimensional Password generation technique

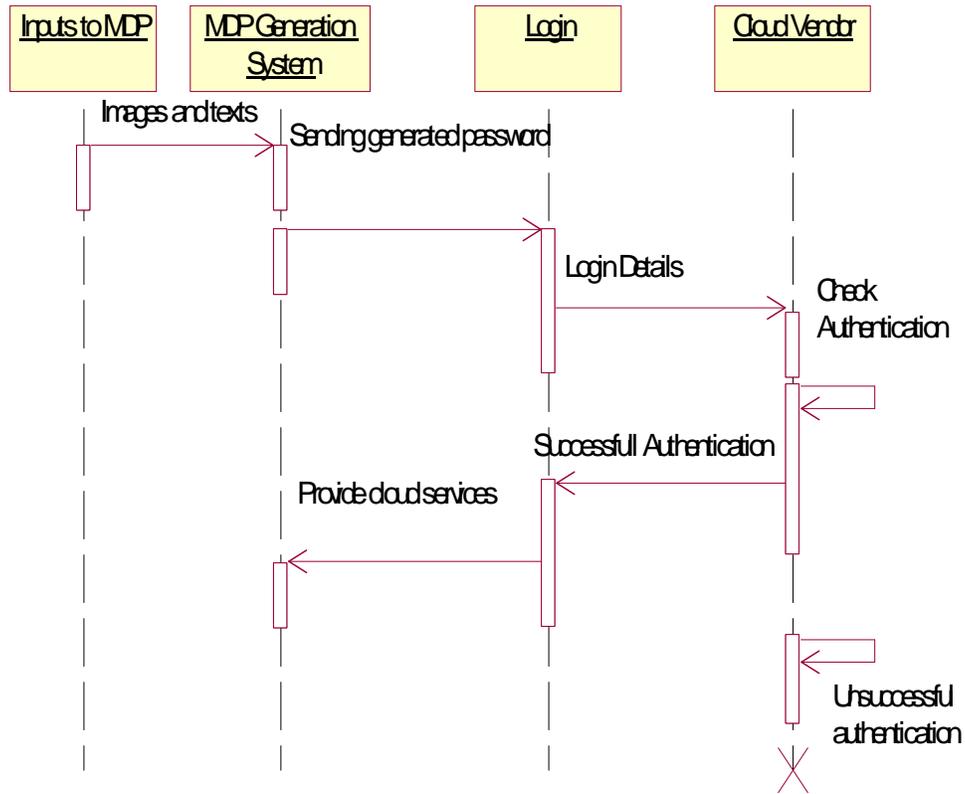

Fig. 3: Sequence diagram to generate multi-dimensional password

## 3. DETAILED DESIGN OF MULTI-LEVEL AUTHENTICATION PROCESS

This section presents algorithm and typical screenshots of multi-dimensional password generation technique. Fig 4 and 5 presents the DFD at level 0 and at level 1 for Multi-dimensional password generation system. Data flow diagram

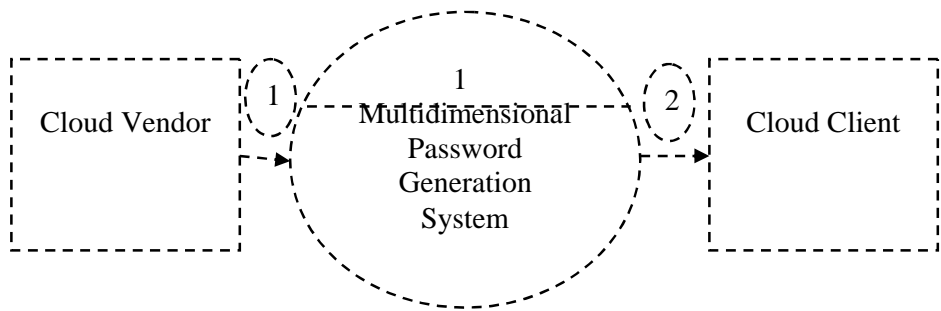

Fig. 4: DFD Level 0: Overview of multi-dimensional password generation System

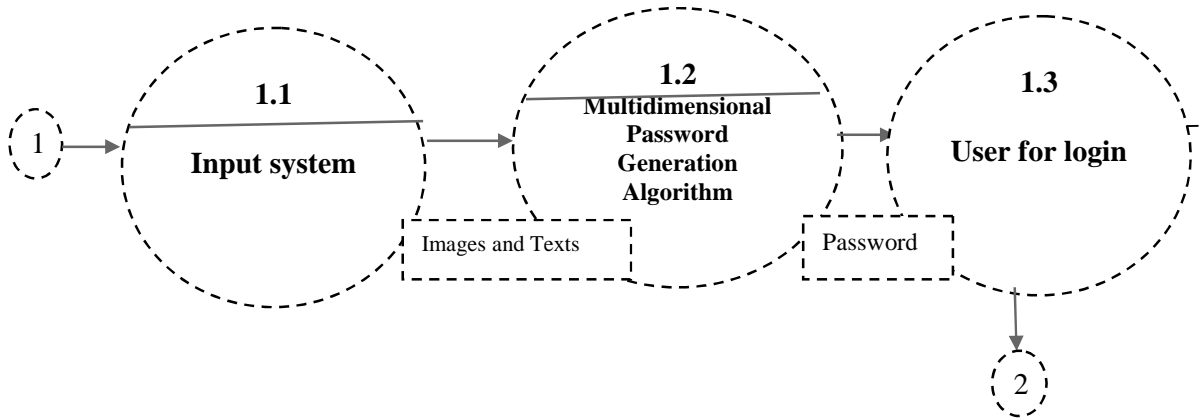

Fig. 5: DFD Level 1 – Multi-dimensional password generation process

The following section depicts the algorithm of complete Multi-dimensional password generation technique.

### 3.1 Algorithm: Multi-Dimensional Password Generation Technique

Algorithm: Generate_Multidimensional_Password ()
Step 1: Read input values
    Read company name, company logo
   Read company signature and company ID
Step 2: Group Images and texts (if any) separately
Step 3: Extract image feature
Step 4: Combine image features with input texts in a pre-defined sequence
Step 5: Send generated password

Step 6: Finish

### 3.2 Typical Graphical User Interfaces

The screenshot shown below presents the typical user interfaces of Multi-dimensional password generation system.

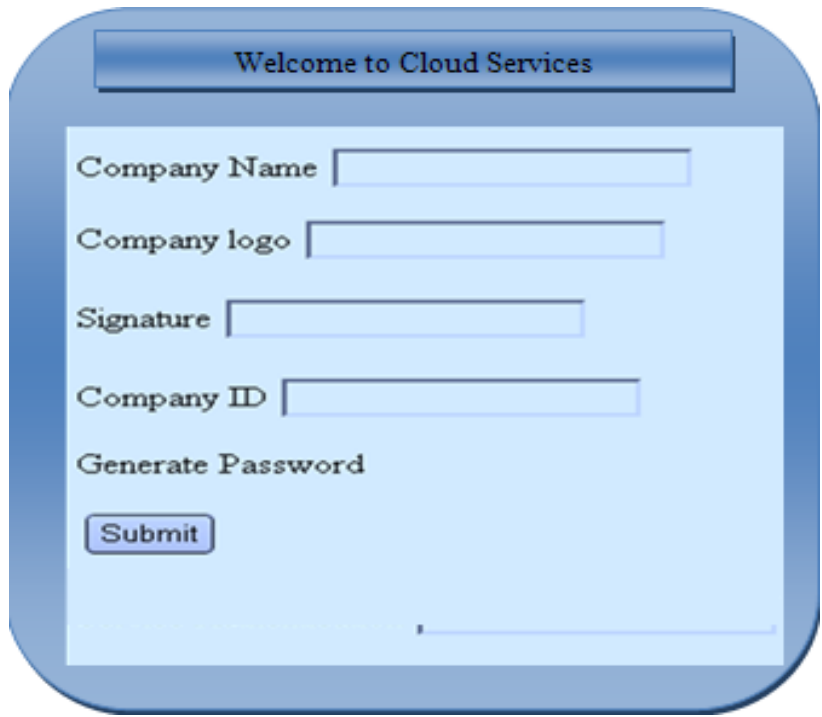

Fig. 6: Typical screenshot of multi-dimensional password authentication

### 3.3. Probability of Breaking Multi-dimensional Authentication

As shown in fig.2, there are 'n' input images for authentication system and each image has got multiple- options. In order to authenticate the password, it is required to send input option in a particular sequence while the following gives a probabilistic model of such an authentication system.

Let 'n' be the number of inputs where the ith input has Ni option. It is assumed that the option in any input selected for the authentication is one at a time. Considering this aspect, we have $n*N_1*N_2…*N_n$ possible options. Therefore, the probability of hacking correct password is $1/n*N_1*N_2*N_3…*N_n => 1/n*N^n$
Assuming all images that we have the same number of option $N_1=N_2=N_3=N$

Given N=100 and number of input n=3, 4, 5 the probability of hacking decreases as number of input increases which is very obvious and shown in figure 7. Given n=3 and number of attempts N=100,200,300,400,500 the probability of hacking decreases as the number of attempts increases which is graphically plotted in figure 8. In both the graphs, Y axis values are logarithmic.

Below figure 7 represents the results while we taken multidimensional password (MDP) generation system with number of attempt of hacking is 100 and increases the number of inputs from 3 to 5.

Table 1: Repesents the MDP with 100 attempts for 3 to 5 input

| Attempts = 100 ||
|---|---|
| n= Number of Input | H= Probability of Hacking |
| 3 | 3.33333E-07 |
| 4 | 2.5E-09 |
| 5 | 2E-11 |

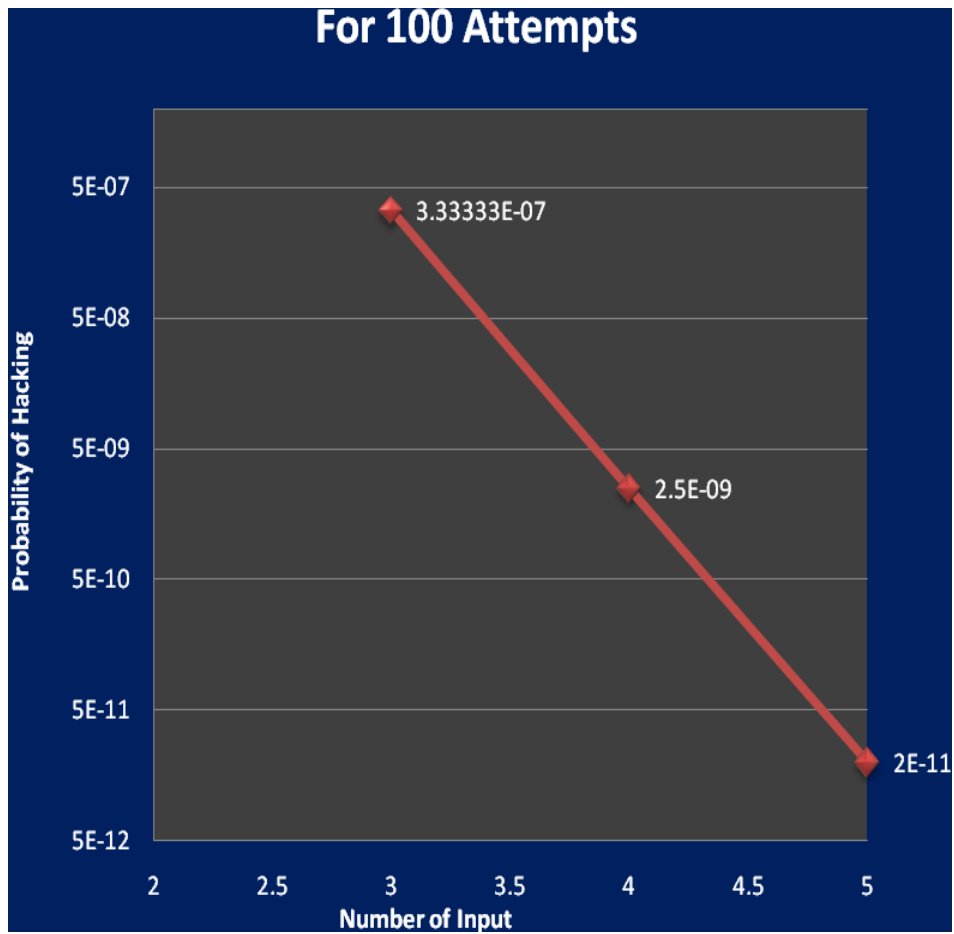

Fig 7: Shown the probability of hacking for given number of inputs

Below figure 7 represents the results while we taken a multidimensional password generation system with number of input is 3 and number of attempts for hacking increases from 100 to 500.

Table 2: Represents the MDP with 3 input and attempts from 100 to 500

| Input = 3 | |
|---|---|
| n= Number of attempts | H= Probability of Hacking |
| 100 | 3.33333E-07 |
| 200 | 4.16667E-08 |
| 300 | 1.23457E-08 |
| 400 | 5.20833E-09 |
| 500 | 2.66667E-09 |

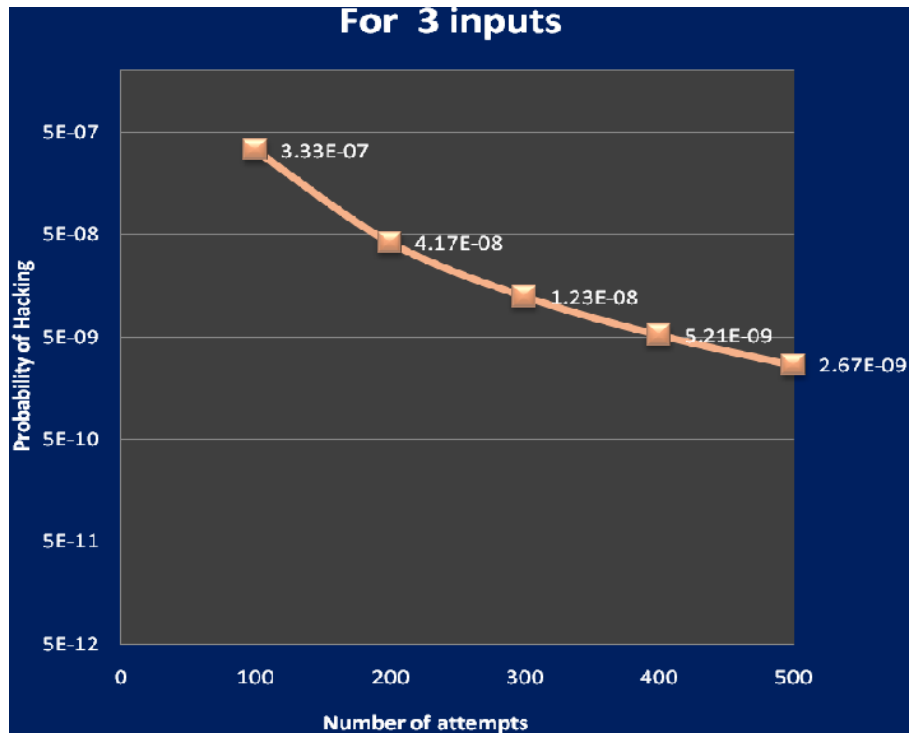

Fig 8: shown the probability of hacking for given number of attempts

From the above graphs, it is very clear that security level i.e. reduction of probability of hacking, improves drastically with increase in number of dimension of input. So it is advised to go multidimensional password authentication scheme. However, based on the level of security requirements one can decide the number of dimension for the input. Therefore, we conclude that by using multi-dimensional password generation technique, we can improve the security of folder

## 4. CONCLUSION AND FUTURE ENHANCEMENT

Cloud computing provides variety of internet based on demand services like software, hardware, server, infrastructure and data storage. To provide secured services to intended customer, we have used multi-dimensional password generation technique. The multi-dimensional password gets generated by considering many aspects and inputs such as, logos, images, textual information's and signatures etc. By doing so, the probability of brute force attack for breaking the password can be reduced to a large extent. Our future work concentrates on the work for intruder detection technique.

### ACKNOWLEDGMENTS

Our sincere thanks to Prof. K N B Murthy Principal & Prof. Shylaja S S, Head Department of Information Science and Engineering, PESIT, Bangalore, for their constant encouragement.

### REFERENCES

[1] Center Bo Wang, HongYu Xing "The Application of Cloud Computing in Education Informatization, Modern Educational Tech..." Computer Science and Service System (CSSS), 2011 International Conference on IEEE, 27-29 June 2011, 978-1-4244-9762-1, pp 2673 - 2676
[2] NIST Definition http://www.au.af.mil/au/awc/awcgate/nist/cloud-def-v15.doc
[3] Cloud Computing services & comparisons http://www.thbs.com/pdfs/Comparison%20of%20Cloud%20computing%20services.pdf
[4] Safiriyu Eludiora1, Olatunde Abiona2,Ayodeji Oluwatope1, Adeniran Oluwaranti1, Clement Onime3,Lawrence Kehinde "A User Identity Management Protocol for Cloud Computing Paradigm" appeared in Int. J. Communications, Network and System Sciences, 2011, 4, 152-163
[5] X. Suo, Y. Zhu, G. S. Owen, "Graphical passwords: A survey," in Proc. 21st Annual Computer Security Application. Conf. Dec. 5–9, 2005, pp. 463–472.
[6] S. Wiedenbeck, J. Waters, J.-C. Birget, A. Brodskiy, and N. Memon, "Authentication using graphical passwords: Basic results," in Proc. Human-Comput. Interaction Int., Las Vegas, NV, Jul. 25–27, 2005.
[7] Fawaz A. Alsulaiman and Abdulmotaleb El Saddik,"Three-Dimensional Password for More Secure Authentication", Instrumentation and Measurement, IEEE Transactions , 03 April 2008, 57 , Issue:9 , 1929 – 1938

### Authors

F. A. Dinesha H A was working in VMware pvt India ltd. Now he is with the PES Institute of Technology. He has designated as Assistant Professor ISE & Research Scientist CORI. Address: 100ft Ring Road, BSK III Stage, Bangalore -560085. Karnataka India (phone: +91-9945870006; FAX: 08026720886, email:sridini@gmail.com)

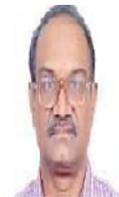

S. B. Dr V. K Agrawal was worked in ISRO and GM. Now he is with the PES Institute of Technology. He has designated as Professor ISE & Director CORI. Address: PESIT 100ft Ring Road, BSK III Stage, Bangalore -560085. Karnataka India (Ph: 080-26720783 FAX: 08026720886, vk.agrawal@pes.edu)